# Polymer-Iron Oxide Hybrid Films for Controlling Electrokinetic Properties

*Austin Dick[1], Xiao Tong[2], Kim Kisslinger[2], Carlos E. Colosqui[1,3,4]\* and Gregory Doerk[2]\**

[1] Department of Mechanical Engineering, Stony Brook University, Stony Brook, NY 11794, United States of America
[2] Center for Functional Nanomaterials, Brookhaven National Laboratory, Upton, NY 11973, United States of America
[3] Department of Applied Mathematics & Statistics, Stony Brook University, Stony Brook, NY 11794, United States of America
[4] Institute of Sustainability, Electrification and Energy, Stony Brook University, Stony Brook, NY 11794, United States of America

**ABSTRACT**
Electrokinetic phenomena at polymer-water interfaces are central to technologies for water purification, ion separations, and energy conversion, yet the ability to systematically control polymer surface charge and associated electrokinetic processes remains limited. Here, we demonstrate a simple liquid-phase infiltration (LPI) method to synthesize polymer–metal oxide hybrid films with controllable interfacial properties. Hydroxy-terminated poly(2-vinylpyridine) (P2VP-OH) brushes grafted to silicon substrates were infiltrated with iron nitrate from ethanolic solution, followed by low-temperature thermal treatment to convert the infiltrated precursor into iron oxide. Spectroscopic ellipsometry, X-ray photoelectron spectroscopy, and thermogravimetric analysis confirmed oxide incorporation and hybrid film formation without polymer degradation. Electrokinetic flow characterization reveals that the hybrid films acquire the electrokinetic properties of the infiltrated oxide, with concentration-dependent streaming potentials and surface conductivities closely matching those of pure iron oxide films. These results establish metal oxide infiltration as a scalable and low-cost strategy for controlling interfacial charge in polymer surfaces. The approach introduces new materials and design parameters for tailoring ion selectivity, transport, and energy conversion, with broad implications for the development of advanced membranes, electrokinetic harvesting devices, and polymer-supported oxide electrodes.

## 1. INTRODUCTION
Control of electrokinetic phenomena at polymer-water interfaces is critical for securing and sustaining future supplies of energy, clean water, and critical resources.[1] Indeed, controlling water interactions at complex, nanoscale solid–aqueous interfaces is integral to the design of high-performance electrodes and catalysts,[2–4] neuromorphic iontronics,[5–7] and separation membranes.[8,9] Electrostatic interactions at water-solid interfaces directly influence surface energy, nanoparticle adsorption and fouling;[10] in the membrane space, these charge interactions are a key determinant of solute selectivity[9] and underpin emerging energy harvesting technologies.[11,12]

For engineering applications, polymers are often favored for low cost and ease of processability, but tailoring polymer surface charge can require not only adoption of alternative polymer chemistries, but also the redevelopment of processes (e.g., casting) used to incorporate them into functional materials. This has motivated the use of atomic layer deposition (ALD) as a general post-fabrication method to modify polymer interfacial properties by depositing very thin metal oxide or other inorganic layers onto surfaces.[13] The application of ALD benefits from its precise control of film thickness, the conformability of deposition through nanoscale porous networks, and the wide and growing library of materials available that can be deposited. Recognizing the importance of aqueous interfacial properties of ALD metal oxide films, Darling et al. have recently published a survey of water contact angles and zeta potentials for 17 metal oxides deposited by ALD.[14]

While ALD deposits thin films onto the surface of polymers by design, other emerging methods for modifying polymer properties after film deposition, known broadly as infiltration synthesis methods, deposit inorganic material both at the outer surface and within the bulk of the polymer, effectively creating a hybrid polymer-inorganic material.[15] As an exemplary case, vapor phase infiltration (VPI), also known as selective infiltration synthesis (SIS), relies on the same chemical precursors and instruments as ALD, but with longer vapor exposure and purge steps for vapor sorption, diffusion, entrapment, and reaction within the polymer matrix.[16,17] VPI-prepared hybrid materials can exhibit enhanced thermal and solvent stability,[18,19] mechanical resilience,[20] triboelectric power,[21] electrical conductivity,[22] and lithographic contrast and etch resistance.[23] VPI may also be used to shrink membrane pore size[24,25] and enable organic solvent separations.[19] Despite the unique advantages afforded by VPI-synthesized polymer-metal oxides, however, their aqueous interfacial properties such as surface charge, are not well understood.

Other infiltration synthesis methods use liquid solvents as the medium to transport metal salts into polymer films, where these salts bind with receptive polymer moieties via electrostatic interaction or complexation.[26–31] In these liquid phase infiltration (LPI) methods, the liquid must be a solvent for both the salt and the polymer, wherein the solvency for the latter may be substantially increased by the addition of acid[28,31] or heat.[32] Block copolymers[28,30,31] and surface grafted polymer brushes[29] have been investigated in past research using LPI as the presence of a insoluble polymer block or the covalent attachment to a substrate, respectively, prevent dissolution of the polymer film. Like VPI, LPI can be used to infiltrate polymers with a broad palette of inorganic species. Moreover, many salts are relatively inexpensive compared to metalorganic ALD precursors used in VPI and the simplicity of LPI lends itself to continuous large-area processing without vacuum equipment. To date, however, all published reports of LPI that we are aware of remove the polymer through oxygen reactive ion etching, ultraviolet ozone degradation, or similar treatment, generating a nanostructured metal or metal oxide film, but not a polymer-inorganic hybrid material as in the case of VPI.

In this work, we develop a simple approach to synthesizing polymer-metal oxide hybrid films using LPI and take initial steps towards understanding the aqueous interfacial properties of hybrid inorganic-polymer films created through infiltration synthesis. We focus on films comprising hydoxy-terminated poly(2-vinylpyridine) (P2VP-OH) "brushes" grafted to oxidized silicon surfaces and infiltrated with iron nitrate or aluminum nitrate from ethanolic solutions. Leveraging the low decomposition temperatures of these salts in comparison to the P2VP-OH brush, we demonstrate conversion of the infiltrated polymer to a polymer metal-oxide hybrid by low temperature thermal annealing. Electrokinetic measurements of the synthesized hybrid films over a wide range of ionic strengths demonstrated that the hybrid polymer-oxide surfaces largely adopt the electrokinetic properties of the metal oxide, with coupling coefficients closely matching those of pure metal oxide surfaces. These results indicate that metal oxide infiltration provides an effective means to control electrokinetic properties in polymer films and surfaces.

## 2. EXPERIMENTAL METHODS

**2.1 Materials.** Hydroxy-terminated P2VP (P2VP-OH, Mn = 6.2 kg/mol) was obtained from Polymer Source Inc. Ethanol, 2-methoxyethanol, iron nitrate nonahydrate and aluminum nitrate nonahydrate were obtained from Sigma-Aldrich. Silicon wafers (~500 μm thick, ⟨100⟩ orientation, N-type, resistivity range of 0–100 ohm-cm) were obtained from University Wafer.

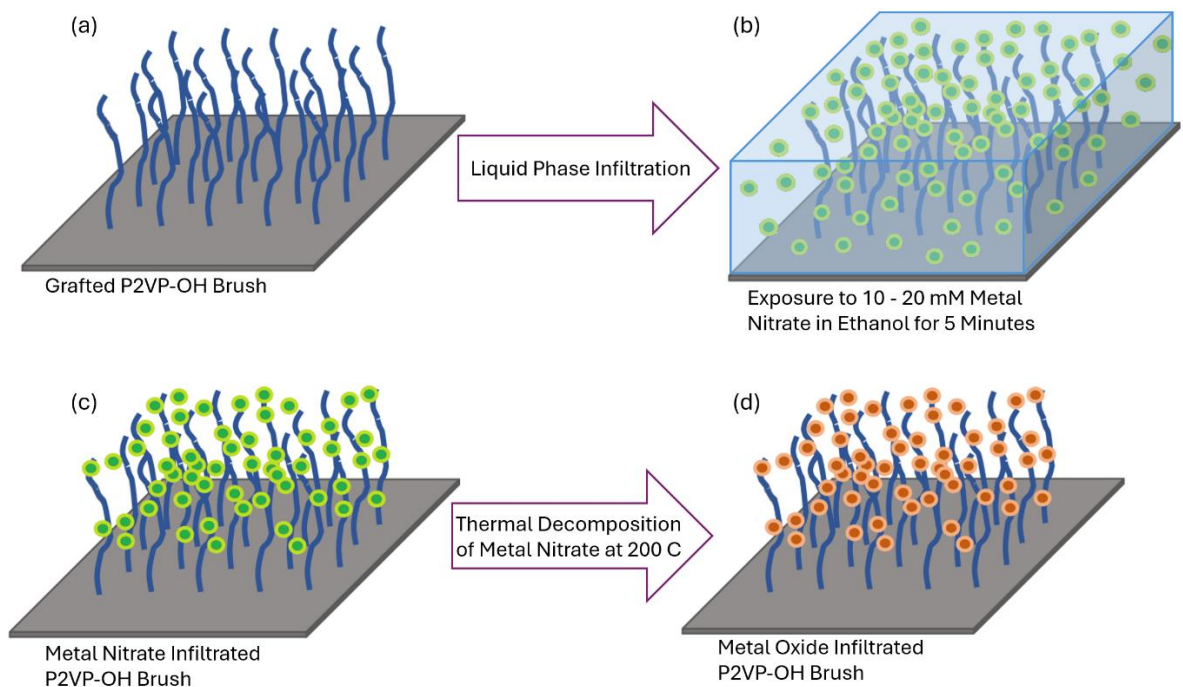

**Figure 1.** Polymer–metal oxide hybridization via LPI. (a) P2VP–OH brushes are grafted onto a silica substrate. (b) The polymer brush is exposed to a metal nitrate solution during LPI, enabling metal nitrate infiltration. (c) Metal nitrate species infiltrated within the polymer film are subsequently converted to a metal oxide through low-temperature thermal decomposition. (d) Hybrid polymer -metal oxide film.

**2.2 Hybrid Film Synthesis.** Covalent grafting of P2VP-OH to silicon wafer surfaces was accomplished as reported previously.[33,34] Briefly, substrates were cleaned by exposure to oxygen plasma for 60 s at 20 W RF power and 100 mTorr total pressure in a reactive ion etcher (Nordson March CS-1701 RIE). Thin films of P2VP-OH were then spin-cast onto substrates from solution in 2-methoxyethanol (1% w/w) at 1500 rpm for 30 s. The films were then baked at 250 °C for 5 minutes under nitrogen purging (Apogee 200 Bake Plate) to graft the hydroxy-terminated P2VP to hydroxy-terminated oxidized silicon via a dehydration reaction. Ungrafted polymer chains were subsequently removed by rinsing with neat 2-methoxyethanol on a spin coater and drying at 3000 rpm for 30s.

The LPI process (Fig. 1) is performed in the case of iron, by immersing brush-grafted substrates in 10-20 mM ethanolic solutions of metal nitrate for 5 minutes at ambient temperature. Samples were then rinsed for 5 seconds under a stream of neat ethanol. Infiltrated metal nitrates were then converted to oxides within the polymer film by baking the samples at 200 °C in air for 10 minutes. For aluminum the same procedure is utilized with a 20 mM ethanolic solution of metal nitrate. Pure metal oxide films (i.e., non-infiltrated) were prepared by spin coating metal nitrate films from ethanolic solutions at 1100 rpm and converting them to metal oxide by baking at 200 °C in air for 10 minutes.

**2.3 Film Characterization.** Measurement of the thickness and refractive indices of the P2VP brushes after grafting, nitrate infiltration and nitrate conversion was performed using an M-2000 (J.A. Woollam) spectroscopic ellipsometer at a 65° incidence angle over a range of wavelengths from 210 nm to 1700 nm. The films were modeled using the Cauchy dispersion model. X-ray photoelectron spectroscopy (XPS) experiments were performed in an ultrahigh vacuum (UHV) system with a base pressure below $3 \times 10^{-9}$ Torr. The system was equipped with a hemispherical electron energy analyzer (SPECS, PHOIBOS 100) and a twin-anode X-ray source (SPECS, XR50). Al Kα radiation (1486.7 eV) was used, generated at 15 kV accelerating voltage and 20 mA emission current. The angle between the analyzer and the X-ray source was 45°, and photoelectrons were collected along the surface normal of the sample. XPS spectra were analyzed using CasaXPS software, with binding energies calibrated to and intensity peaks normalized by the elemental Si 2p peak at 99.3 eV.

**2.4 Thermal analysis.** Thermal behavior of the materials, including decomposition temperatures, were characterized by thermogravimetric analysis (TGA; Mettler Toledo TGA/DSC 3+) under a continuous temperature increase from 25 to 500 °C at 20 °C/min with 50 mL/min of oxygen flow.

**2.5 Characterization of Electrokinetic Properties.** The electrokinetic coupling coefficient, defined as the ratio of streaming potential to pressure head, and the electrical conductivity, comprising bulk, surface, and excess components, was measured for all surface samples using a custom built flow cell specifically designed for the analysis of planar substrates.[35] The flow cell (cf. Fig. 2) consists of two inlet/outlet reservoirs connected by a slit microchannel of height $H = 20$ μm, length $L = 3.81$ cm, and width $W = 2.54$ cm, these dimensions enable the precise measurement of voltage differences, pressure differentials, and electrical conductance across a wide range of electrolyte concentrations.

The surface samples studied were integrated into the top and bottom surfaces of a microfluidic slit channel within the experimental flow cell. DI water and water-NaCl electrolyte solutions with different concentrations ($C = 0.1, 1.0, 10$ & $50$ mM) were injected through the inlet reservoir at controlled volumetric flow rates ($Q = 0.9, 1.0,$ & $1.1$ mL/min) using a syringe pump (Chemyx Fusion 200). The solution remained at nearly constant pH = $5.5 \pm 0.1$ for all experiments. Simultaneous measurements of pressure and electrical potential differences across the slit microchannel were collected over time periods of 400 seconds to evaluate steady electrokinetic behavior. Electrical potential measurements were obtained using a Keithley DMM-6500 digital multimeter, which offers a high input impedance of 10 GΩ for voltages below 12 V, enabling precise measurement of voltages at low electrolyte concentrations. Pressure measurements were recorded using an Idex I2C PS200 pressure sensor with a resolution of 2 psi. As illustrated in Figure 2b, the system attains steady-state conditions within 120 seconds. Upon reaching steady state, the time-averaged pressure head (Δp) across the channel and voltage difference (ΔV) between the electrodes, separated by a distance $L_v = 2.54$ cm are measured for each flow rate (cf. Fig 2b). The relative uncertainties associated with these measurements remained below 5% for all reported cases. Under steady-state conditions, the time-averaged pressure head Δp exhibits a linear increase with respect to the flow rate, in agreement with the analytical predictions for Poiseuille flow. This observation corroborates a nominal channel height of $H = 20 \pm 8\%$ μm. Furthermore, the experimentally determined average voltage difference $\Delta V$ is found to be linearly proportional to $\Delta p$. Consequently, the electrokinetic coupling coefficient, defined as the streaming voltage-to-pressure slope $dV/dp = (\Delta V/\Delta p) \times (L/L_v)$, is obtained via linear regression analysis of experimental data (Fig 2c). Electrical resistance measurements across the flow cells containing the studied surface samples were performed independently, under no-flow conditions, to determine the intrinsic electrical conductivity of the studied surfaces. The time-averaged resistance ($R_E$) measured between the electrodes at different NaCl concentrations was converted to the effective electrical conductivity, $\kappa_E = R_E^{-1} L_v/(WH)$, across the channel test section. These conductivity measurements were conducted over the same range of NaCl concentrations ($C = 0.01–50$ mM) for the pure P2VP–OH brush, the standalone iron oxide film, and the hybrid polymer–iron oxide surface to enable direct comparison.

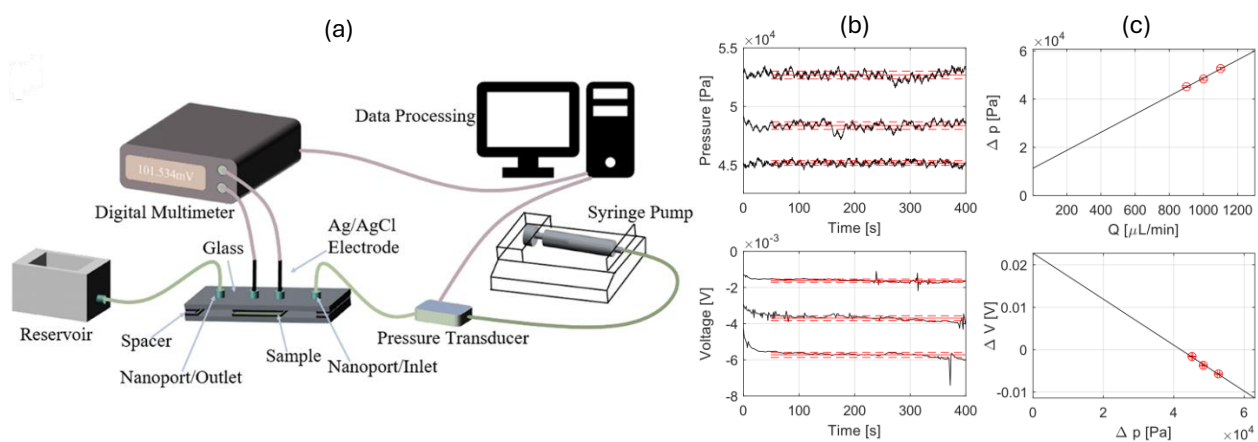

**Figure 2.** Electrokinetic cell experimental setup and example measurements. (a) Schematic of experimental setup (b) Pressure-time and voltage-time traces conducted on the hybrid polymer-iron oxide film using 1.0 mM NaCl solution. (c) Linear regressions of pressure vs flowrate as well as voltage vs pressure conducted on the hybrid polymer-iron oxide film using 1.0 mM NaCl solution.

## 3. RESULTS AND DISCUSSION

**3.1 Polymer Surface Hybridization.** The general process flow for hybrid film synthesis is depicted schematically in Figure 1. Silicon substrates grafted with P2VP-OH monolayer brushes are immersed in 10-20 mM solutions of a metal nitrate in ethanol. As a solvent for the nitrates and P2VP, ethanol swells the brush, ensuring efficient nitrate diffusion within the brush layer. Metal nitrate cations then bind to the pyridine groups via coordination.[36] After drying, metal nitrates then decompose to metal oxides within the polymer brush layer upon heating at 200 °C, forming a polymer-metal oxide hybrid film. The low decomposition temperatures of metal nitrates[37] are critically important for the efficacy of the hybrid film synthesis strategy demonstrated here, as it enables metal nitrate-to-oxide conversion with minimal impact on the polymer within which it resides. As an exemplar, we focus on $Fe(NO_3)_3 \cdot 9H_2O$ as precursor to form iron oxide ($FeO_x$) based on its potential use as a catalytic material for water splitting[38] and the degradation of pharmaceutical water pollutants.[39] As shown by thermogravimetric analysis (TGA) performed in an oxygen atmosphere (Figure 3a), $Fe(NO_3)_3 \cdot 9H_2O$ begins to decompose at around ~100 °C, with complete decomposition by ~170 °C. In contrast, aside from a modest weight loss around ~100 °C that can be attributed to water evaporation, the P2VP-OH brush does not decompose until temperatures reach above 300 °C. The large difference in decomposition temperatures between the Fe(NO$_3$)$_3$•9H$_2$0 and the P2VP-OH implies a substantial window for thermal conversion to iron oxide without affecting the polymer. For comparison, iron chlorides decomposes to ferric oxide at temperatures in excess of 280 °C depending on the hydration state,[40] leaving little room for metal oxide conversion without polymer degradation.

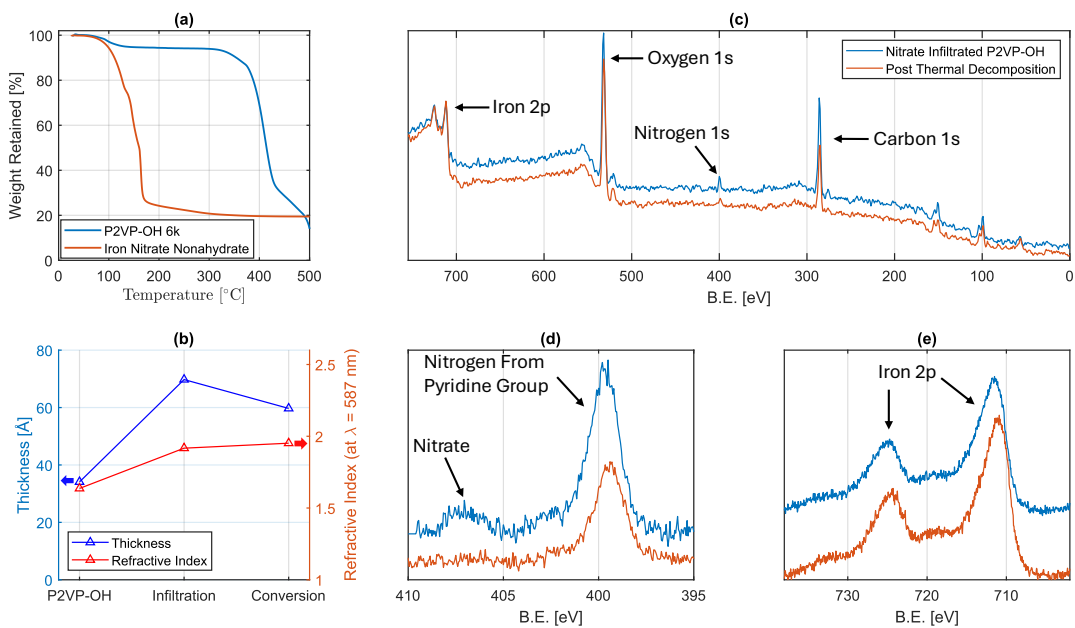

**Figure 3.** Hybrid polymer surface characterization. (a) Thermogravimetric analysis of P2VP-OH and iron(III) nitrate nonahydrate. (b) Ellipsometric thickness and refractive index measurements performed after each step of the hybrid film synthesis process. (c–e) XPS analysis of P2VP-OH films infiltrated with iron(III) nitrate, before and after thermal conversion to iron oxide: (c) wide-scan XPS survey spectrum, (d) high-resolution N 1s region, and (e) high-resolution Fe 2p region.

Measurements of the film thickness and refractive index of the polymer brush film using spectroscopic ellipsometry after each step provide a means to estimate the efficacy of iron infiltration and conversion (Fig. 3b). The film thickness of the grafted P2VP-OH brush after grafting is ~3.5 nm, similar to what has been reported previously for P2VP-OH with the same molecular weight,[29] while the refractive index is ~1.65, which is comparable to that for similar polymers.[34] The film thickness increases to ~7 nm after Fe(NO$_3$)$_3$•9H$_2$0 infiltration while the refractive index increases to more than 1.9, indicating a significant volume of Fe(NO$_3$)$_3$•9H$_2$0 incorporated into the brush film. After thermal treatment at 200 °C to promote conversion, the film thickness decreases to ~6 nm while the refractive index increases to ~2. Together these imply a densification of the Fe(NO$_3$)$_3$•9H$_2$0 due to the loss of volatile species associated with decomposition.

X-ray photoelectron spectroscopy (Fig. 3c-e) was used to characterize the conversion of iron nitrate to oxide within the polymer brush film. The survey spectrum (Fig. 3c) shows that oxygen and iron peak intensities increase relative to that of carbon, suggesting a change in the density of iron in the sample, in line with the ellipsometry results. Meanwhile, a peak in the N 1s region at

~407 eV associated with nitrate ligands that are present after infiltration disappears upon thermal treatment (Fig. 3d), while the iron 2p peaks remain after the same step (Fig. 3e). After thermal treatment, the Fe 2p$_{3/2}$ peak shifts by ~0.7 eV compared with Fe(NO$_3$)$_3$·9H$_2$O to a binding energy of 711 eV and the shake-up satellite around ~719 eV becomes slightly stronger. This combination indicates that Fe has transformed from a nitrate-coordinated Fe$^{3+}$ environment into a predominantly Fe$^{3+}$ oxide environment similar to Fe$_2$O$_3$.[41,42] The lower binding energy reflects loss of the strongly electron-withdrawing nitrate ligands, while the enhanced satellite intensity is characteristic of Fe$^{3+}$ in oxides. Taken together, these results strongly support the assessment that the thermal annealing at 200 °C converts the Fe(NO$_3$)$_3$•9H$_2$0 to FeO$_x$.

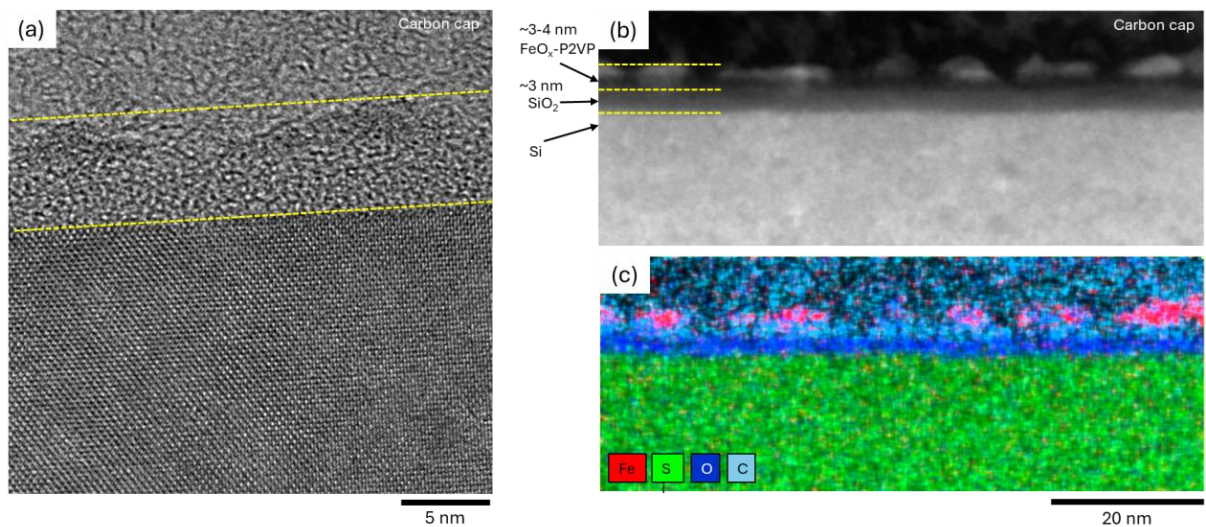

**Figure 4.** Surface characterization. (a) High resolution bright-field TEM image of the P2VP-iron oxide hybrid film on top of a native silicon oxide layer. The dashed yellow line marks the approximate thickness of the region containing both. (b) HAADF-STEM and (c) color-coded EDS elemental map of the infiltrated and converted film.

Transmission electron microscopy (TEM) was used to characterize structural properties of the hybrid film after conversion. A high resolution bright-field TEM image (Fig. 4a) shows a ~6 nm thick region comprising the native silicon oxide (SiO$_2$) and the overlying hybrid FeO$_x$-P2VP film, demarcated with a dashed yellow line. While slightly darker regions near the top of the layer may be tentatively attributed to FeO$_x$, the SiO$_2$, hybrid film and carbon cap are all amorphous and effectively indistinguishable. High-angle annular dark field scanning TEM (HAADF-STEM, (Fig. 4b) and energy dispersive spectroscopy (EDS) mapping (Fig. 4c) were therefore used to characterize the elemental distribution. A 3-4 nm thick hybrid FeO$_x$-P2VP film, evident from regions of higher atomic number contrast (i.e., brighter in HAADF-STEM), lies on top of an ~3 nm thick oxygen-rich native SiO$_2$ layer. The presence of iron is confirmed by the EDS map (Fig. 4c). It is evident that the iron is not uniformly distributed laterally, which we expect can be remedied by further process optimization. The HAADF-STEM image and EDS map also reveal a ~1-2 nm gap between FeO$_x$ and the native SiO$_2$ layer, indicating that a majority of the iron oxide resides closer to the surface of the FeO$_x$-P2VP hybrid film, similar to what has been observed previously in copper nitrate infiltration.[43] Additional clarification of the elemental distribution is provided by individual EDS elemental maps for silicon, carbon, oxygen, and iron in the Supporting Information (Fig. S1).

This general approach to hybrid polymer-metal oxide film synthesis can readily be applied to other metal nitrates. As an example, we used the same method to create hybrid films with aluminum oxide, a metal oxide of interest for its high dielectric constant[44] and high positive zeta potential in mildly acidic conditions.[14] For infiltration, we employed $Al(NO_3)_3 \cdot 9H_2O$ at a concentration of 20 mM in ethanol and decomposed the nitrate to an oxide within the polymer by baking 10 minutes in air at 200 °C, well above the aluminum nitrate decomposition temperature of ~160 °C.[37] Successful infiltration and conversion was confirmed by XPS through the presence of a persistent peak at ~75-76 eV in the aluminum 2p region (which shifts to lower binding energy by ~0.4 eV due to the loss of electron-withdrawing nitrate ligands) and the disappearance of nitrate and N-O peaks[43,45] from the N 1s region (Fig. S2). Unlike the $FeO_x$ films derived from $Fe(NO_3)_3 \cdot 9H_2O$ thermal conversion, however, these nitrate-derived aluminum oxide films are unstable in salt solutions as shown by XPS spectra acquired before and after "aging" for 7 days in 10 mM aqueous NaCl solutions (Fig. S3). This result, which is consistent with findings previously reported by Willis et al. for ALD-deposited amorphous alumina,[46] prevented characterization of the electrokinetic behavior of hybrid polymer-aluminum oxide surfaces.

**3.2 Electrokinetic Properties.** Experimental analysis is performed for the three studied surface samples (1) the pure polymer brush surface, Hydroxy-terminated P2VP-OH grafted on silicon, (2) the pure metal oxide surface, standalone $FeO_x$ film (amorphous, predominantly $Fe_2O_3$-like phase), and (3) the hybrid polymer-oxide surface, with P2VP-OH brush infiltrated with iron nitrate, thermally converted to iron oxide ($FeO_x$) within the polymer matrix. The electrokinetic behavior of the studied surfaces is characterized by their surface charge $\sigma$ and zeta potential $\zeta$, quantities that are regarded as material properties dependent on the electrolyte solution concentration $C$ and $pH$, and, in particular, on the electric double layer (EDL) thickness characterized by the Debye length $\lambda_D \propto C^{-1/2}$. These physical parameters determine the electrokinetic coupling coefficient $\Delta V/\Delta P$ and the effective electrical conductivity $\kappa_E$, which are directly measured using the flow cell and apparatus described in Sec. 2.5 and quantitatively described by the model presented in this section.

*3.2.1 Analytical model.* The surface charge density, σ, is determined from the surface potential $\psi_o$ through a conventional two-site acid–base equilibrium model[35,47,48]

$$\sigma(\psi_o) = e\Gamma \frac{2\sinh\left[(\psi_N - \psi_o)e/k_BT\right] \cdot 10^{(pH_0 - pK_-)}}{1 + 2\cosh\left[(\psi_N - \psi_o)e/k_BT\right] \cdot 10^{(pH_0 - pK_-)}}. \quad (1)$$

where $\Gamma$ is the density of chargeable surface sites (nm$^{-2}$), $e$ is the elementary charge, $k_B$ is the Boltzmann constant, and $T$ is the absolute temperature. The model parameters $pK_-$ and $pK_+$ correspond to the acid and base dissociation constants of surface sites, and $pH_0 = (pK_+ + pK_-)/2$ defines the isoelectric point, $\psi_N = -2.3(k_BT/e)(pH - pH_0)$ is the so-called Nernst potential. The diffusive layer potential, right outside the Stern layer, is expressed as

$$\psi_d = \psi_o - \frac{\sigma}{C_s}, \quad (2)$$

where $C_s$ is the specific Stern-layer capacitance. Charge neutrality for a symmetric 1:1 electrolyte (in the thin limit $H \gg \lambda_D$) is enforced through the classical Grahame equation

$$\psi_d = \frac{2k_B T}{e} \operatorname{arcsinh}\left(\frac{\sigma e \lambda_D}{2\varepsilon k_B T}\right), \quad (3)$$

where $\varepsilon$ is the permittivity of the electrolyte solution and $\lambda_D = (\varepsilon k_B T / 2e^2 C N_A)^{1/2}$ is the Debye length ($C$ is the electrolyte molar concentration and $N_A$ the Avogadro number).

For given electrolyte concentration, pH, and set of material parameters $\Gamma, pK_+, pK_-, C_s$ (see Table 1), Eqs. (1)-(3) are solved self-consistently to obtain the surface charge density $\sigma$, surface potential $\psi_o$, and diffuse potential $\psi_d$. Under the experimental conditions considered here, with $H \gg \lambda_D$, the zeta potential can be approximated as $\zeta \approx \psi_d$. The electrokinetic coupling coefficient is then given by the Helmholtz–Smoluchowski relation

$$\frac{dV}{dP} = \frac{\varepsilon}{\mu \kappa_E} \psi_d, \quad (4)$$

where $\mu$ is the dynamic viscosity of the electrolyte solution, and $\psi_d$ is the diffuse-layer potential obtained from the solution of Eqs. (1)-(3).
The effective electrical conductivity of the electrolyte–channel system is measured directly and can be expressed as[35]

$$\kappa_E = \kappa_B + \kappa_{EDL} + \kappa_{ex}, \quad (5)$$

comprising the bulk conductivity $\kappa_B$, EDL conductivity $\kappa_{EDL}$, and excess interfacial conductivity $\kappa_{ex}$, the latter treated as an adjustable model parameter to fit the experimental measurements for $C \to 0$.[35] The bulk conductivity is readily measured prior and following the flow experiments and can be analytically accounted for as $\kappa_B = \Lambda_{NaCl} C + N_A \mu_+ 10^{-pH+3}$, where $\Lambda_{NaCl} = 12.64$ S m$^2$/mol is the molar conductivity of fully dissociated Na+ and Cl- ions, and $\mu_+ = 3.63 \times 10^{-7}$ m2/V s is the proton mobility in water associated with the Grotthuss mechanism. The bulk concentration largely dominates the total conductivity for NaCl concentrations above 1 mM. Using the Debye–Hückel approximation, $\psi(y) \simeq \psi_d e^{-y/\lambda_D}$, for a symmetric 1:1 electrolyte the EDL contribution in Eq. (5) can be estimated as $\kappa_{EDL} \approx (2e^2 \varepsilon C N_A \lambda_D \psi_d^2)/(\mu H k_B T)$. The EDL conductivity ($\kappa_{EDL}$) contributes less than 5% to the total conductivity under the conditions studied. The material parameters used to generate the analytical predictions presented in Figs. 5a–d are listed in Table 1 for the three surfaces analyzed experimentally. A Stern capacitance $C_s = 200$ mF/m$^2$ is employed in all cases. Using Eq (1), the model predicts a range of positive surface charge $\sigma = 0.14$ to $3.7$ mC/m$^2$ for the P2VP-OH surface and negative surface charges $\sigma = -0.27$ to $-2.7$ mC/m$^2$ for the iron oxide, which is consistent with values reported in previous studies.[14,49] The values of the excess conductivities ($\kappa_{ex}$) reported in Table 1 are obtained by fitting the experimental data, and are consistent with charge transport by highly mobile protons within water confined in nanoscale topographic features at the interface.[35]

| Surface sample | Γ [nm$^{-2}$] | pK$_-$ | pK$_0$ [ | κ$_{ex}$ [S/m] |
|---|---|---|---|---|
| Pure polymer surface (P2VP-OH) | 6.5 | 9.3 | 6.4 | 0.006 |
| Pure metal oxide surface (FeO$_x$) | 3.5 | 7.5 | 4.4 | 0.009 |
| Hybrid polymer-oxide (P2VP-OH+ FeO$_x$) | 3.5 | 7.5 | 4.4 | 0.007 |

**Table 1:** Model parameters employed for analytical predictions of electrokinetic properties.

*3.2.2 Experimental analysis.* Electrokinetic measurements are performed for the studied set of NaCl concentrations $C = 0.1, 1.0, 10$ & $50$ mM. Given that the bulk conductivity measured for DI water under the pH conditions of our experiments is 0.11 mS/cm, corresponding to a molar concentration $C = 0.009$ mM NaCl, this equivalent concertation is used when reporting the results for DI water. The electrokinetic coupling coefficient, $dV/dp$, determined from the ratio between steady-state electrical potential $\Delta V$ and pressure differentials $\Delta p$ over the studied flow rates, is reported in Fig. 5a and for the pure polymer, and Fig. 5c for the FeO$_x$ film, and the hybrid polymer-oxide surfaces.

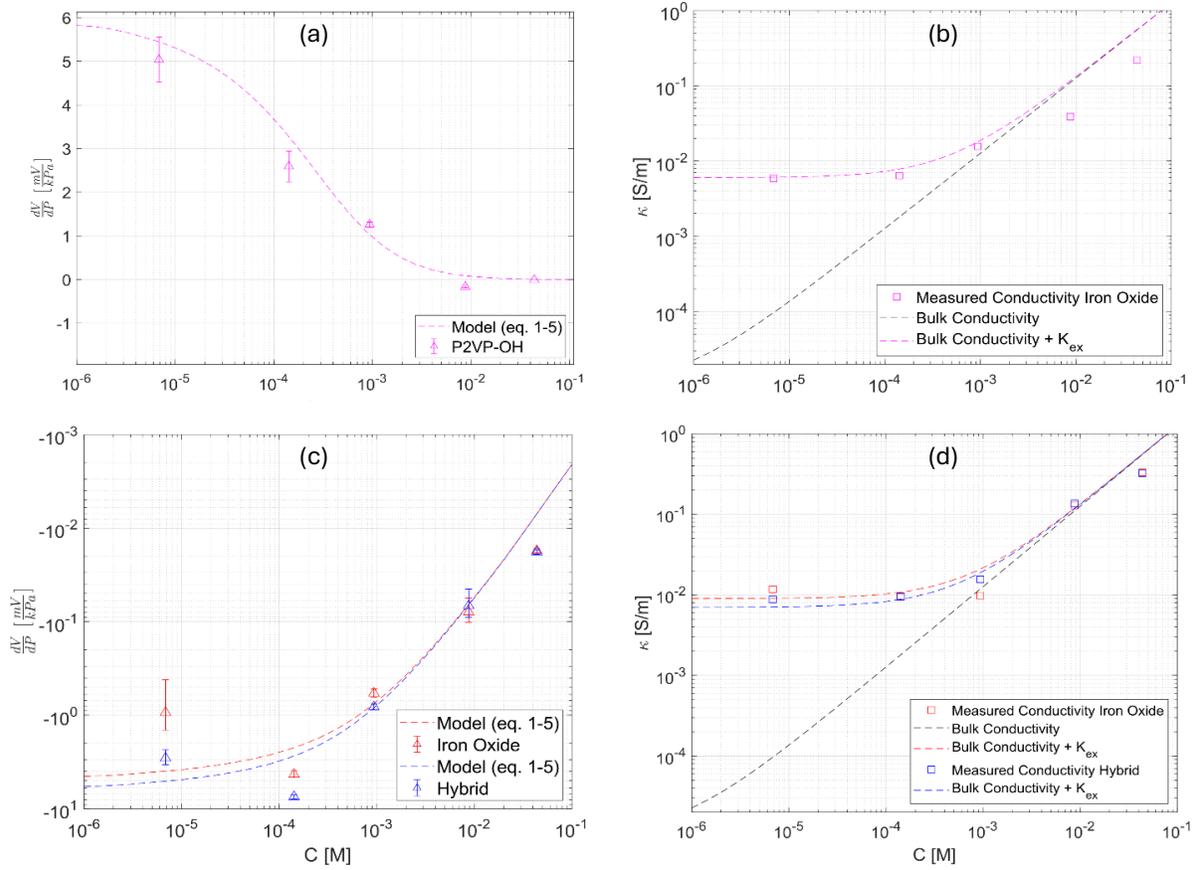

**Figure 5:** Electrokinetic properties of for the three studied surface samples: (1) polymer brush surface, Hydroxy-terminated P2VP-OH grafted on silicon, (2) metal oxide surface, standalone FeO$_x$ film, and (3) hybrid polymer-oxide surface, P2VP-OH brush with iron oxide (FeO$_x$). (a) & (c): Electrokinetic coupling coefficient $dV/dP$ vs. water-NaCl solution concentration: experimental measurements (markers) and analytical predictions (Eq. 4). (b) & (d): Electrical conductivity vs. water-NaCl solution concentration: experimental measurements (markers) and analytical predictions (Eq. 5). Analytical model estimates (Eqs. 1-5) employ the parameters reported in Table 1.

The most striking observation is while that the pure polymer surface exhibits positive electrokinetic coefficients $dV/dP > 0$ for moderate-to-low electrolyte concentrations C ≤ 1 mM and weakly negative ones $dV/dP \lesssim 0$ for C ≥ 10 mM (Fig. 5a), the hybrid polymer-oxide surface displaying a simpler electrokinetic response that closely matches that of the standalone iron oxide surface, with $dV/dP < 0$ over the entire concentration range studied (Fig. 5c). Furthermore, at high NaCl concentrations (C > 1 mM), the pure polymer shows zeta-potential sign reversal and reduced conductivity ($\kappa_E \lesssim \kappa_B$), consistent with substantial NaCl adsorption (Fig. 5a–b); this behavior is not observed for the hybrid or iron oxide surfaces (Fig. 5c–d). The experimental measurements are accounted for by analytical predictions obtained from Eqs. (4)-(5) with the parameters reported in Table 1. The model parameters used to account for the results for the hybrid polymer-oxide surface (cf. Table 1) further confirm that oxide infiltration confers the electrokinetic properties characteristic of the standalone metal oxide. The electrical conductivity $\kappa_E$ experimentally determined for the pure polymer, standalone oxide, and hybrid polymer-iron surfaces is presented along with analytical estimates via Eq. (5) in Fig. 5b and Fig. 5d. As expected, the electrical conductivity reported for all samples follows closely the bulk conductivity estimate $\kappa_E \simeq \kappa_B \propto C$ for NaCl concentrations above 1 mM. For very low concentrations $C \lesssim 0.1$ mM, however, the three surfaces studied report a nearly constant conductivity $\kappa_E \simeq \kappa_{ex}$ prescribed by the corresponding excess conductivity value reported in Table 1. The lowest excess conductivity corresponds to the pure polymer ($\kappa_{ex} \simeq 6.0$ mS/m), while the highest value is observed for the standalone iron oxide ($\kappa_{ex} \simeq 9.0$ mS/m). The hybrid polymer-oxide surface exhibits an intermediate excess conductivity ($\kappa_{ex} \simeq 7.0$ mS/m), consistent with the combined contribution of polymer and oxide phases in electrical contact within the hybrid film. Hence, electrokinetic measurements indicate that the hybrid film exhibits nearly the same zeta potential and surface conductivity as the iron oxide surface over a wide range of electrolyte concentration.

## 4. CONCLUSIONS

This work demonstrates a simple, scalable, and generalizable liquid-phase infiltration (LPI) method for producing polymer-metal oxide hybrid films with tailored electrokinetic properties. Through infiltration of hydroxy-terminated poly(2-vinylpyridine) (P2VP–OH) brushes with iron nitrate followed by low-temperature thermal conversion, we achieved hybridization that effectively conferred the electrokinetic behavior of iron oxide onto the polymer surface. The resulting hybrid films exhibited streaming potentials, electrokinetic zeta potentials, and excess interfacial conductivities closely matching those of standalone FeO$_x$ surfaces, confirming that the oxide phase dominates the interfacial charge and transport characteristics. These findings advance the current understanding of hybrid polymer-metal oxide systems and their role in manipulating electrokinetic behavior under practical conditions.

Beyond establishing this proof of concept, the LPI approach offers distinct practical advantages. The method is highly scalable, compatible with solution processing, and avoids the need for vacuum-based deposition or complex precursors. Because it relies on simple metal nitrate solutions, the same strategy can be readily extended to oxides of other metals such as aluminum, copper, or manganese, enabling the creation of diverse polymer-oxide hybrids with a wide range of electrostatic properties.

Looking forward, refined spatial control over oxide infiltration could enable surfaces patterned with localized regions of opposite charge or designed gradients in surface potential. Achieving uniform or spatially modulated infiltration degrees would provide tunability of the electrokinetic zeta potential over a broad range, from positive to neutral and negative values, opening new pathways for the design of adaptive interfaces. Such hybrid surfaces could serve as model systems for studying charge-regulated transport and as building blocks for next-generation electrokinetic devices, including ion-selective membranes, nanomaterial adsorption and separation platforms, blue-energy harvesters, and iontronic computing devices, such as fluidic memristors, electrochemical neuromorphic elements, or ionic logic gates.

Overall, the presented LPI method establishes a versatile and scalable route to control interfacial charge and electrokinetic behavior in polymer-based materials through selective oxide infiltration, bridging the processability of polymers with the functional properties of inorganic oxides. Expanding the ability to tune the surface charge of polymer films across both positive and negative potentials would greatly enhance the versatility of this approach, extending its relevance to electrokinetic transport, surface catalysis, and the design of advanced polymer-supported oxide electrodes for energy conversion and storage applications.

## ASSOCIATED CONTENT

**Supporting Information**.

The following files are available free of charge.

An HAADF-STEM image and corresponding EDS maps for a hybrid FeOx-P2VP film; XPS spectra of P2VP brushes infiltrated with aluminum nitrate and subsequent conversion; XPS spectra before and after hybrid film aging in aqueous sodium chloride solutions.

## AUTHOR INFORMATION

**Corresponding Authors**

*carlos.colosqui@stonybrook.edu

*gdoerk@bnl.gov

**Author Contributions**

Experimental sample preparation, ellipsometry, and electrokinetic measurements were performed by A.D.; XPS data acquisition and analysis were performed by A.D. and X.T.; STEM imaging was performed by K.K.; results were interpreted by A.D., C.C., and G.S.D.; all authors have contributed to reviewing and editing the manuscript and approve the published version.

## ACKNOWLEDGMENTS

This research used the Materials Synthesis and Characterization, Proximal Probes, and Electron Microscopy facilities at the Center for Functional Nanomaterials (CFN), which is a U.S. Department of Energy (DOE) Office of Science User Facilities, at Brookhaven National Laboratory under Contract No. DE-SC0012704. A.D. was supported in part by the DOE Office of Science Graduate Student Research (SCGSR) Program. C.C. acknowledges support from the National Science Foundation under award CBET-2417797.